\documentclass[12pt]{article}

\usepackage{epsfig,amsfonts,newlfont,rotate,float}

\DeclareMathAlphabet{\mathitbf}{T1}{cmr}{bx}{it}
\newcommand{\mod}{\mbox{ mod}}
\renewcommand{\d}{\mathrm d}
\newcommand{\e}{\mathrm e}

\begin{document}

\title{Measures of critical exponents in the \\
        four dimensional site percolation}
\author{H.~G.~Ballesteros\footnote{\tt hector@lattice.fis.ucm.es},
        L.~A.~Fern\'andez\footnote{\tt laf@lattice.fis.ucm.es},\\
        V.~Mart\'{\i}n-Mayor\footnote{\tt victor@lattice.fis.ucm.es},
        A.~Mu\~noz Sudupe\footnote{\tt sudupe@lattice.fis.ucm.es},\\
 \normalsize \it Departamento de F\'{\i}sica Te\'orica I, 
        Facultad de CC. F\'{\i}sicas,\\  
\normalsize \it Universidad Complutense de Madrid, 28040 Madrid, Spain.\\
\\
        G.~Parisi\footnote{\tt parisi@roma1.infn.it}, and
        J.~J.~Ruiz-Lorenzo\footnote{\tt ruiz@chimera.roma1.infn.it}.\\
\normalsize \it Dipartimento di Fisica and Istituto Nazionale di 
        Fisica Nucleare,\\ 
\normalsize \it Universit\`a di Roma ``La Sapienza'', P.~A.~Moro  2, 
        00185 Roma, Italy.
}

\date{20 December 1996}

\maketitle

\thispagestyle{empty}

\begin{abstract}

Using finite-size scaling methods we measure the thermal and magnetic
exponents of the site percolation in four dimensions, obtaining a value
for the anomalous dimension very different from the results found in
the literature.  We also obtain the leading corrections-to-scaling
exponent and, with great accuracy, the critical density.

\end{abstract}

\vskip 5 mm

Monte Carlo.
Percolation.
Critical exponents.
Finite size scaling.
$\epsilon$-expansion.


\vskip 5 mm

\newpage

\section{\protect\label{S_INT}Introduction}

From the point of view of its definition, the simplest
statistical system is perhaps the percolation. In the case of the site
percolation, we fill the sites of a given lattice with probability
$p$. Then we construct the clusters as sets of contiguous filled sites.

The critical properties of the system can be described in terms of
the clusters. For instance, at the critical percolation the mean
cluster size diverges. We define the percolating cluster as the one
that contains, in the thermodynamical limit, an infinite number of
sites. The strength of this cluster (i.e. the probability of
containing an arbitrary point) is the order parameter of the
transition: it is zero for $p<p_c$, and finite for $p>p_c$
\cite{AHSTA}.

Another interesting model is the bond percolation. In this case we
fill the lattice bonds with a given probability and construct clusters
analogously.  It is believed  that both models
belong to the same Universality Class (share the critical exponents).

It is possible to relate the percolation problem (in the bond version)
with the $q$-states Potts model using the ``Fortuin-Kasteleyn''
representation of the latter. The bond percolation is obtained in the
$q\to 1$ limit \cite{KAFO}.

Moreover it is possible to write down a field theoretical description
of the percolation.  In general, the Potts model is described by means
of a $\phi^3$ theory, where the coefficient of the cubic term is
proportional to $q-2$. For the Ising model ($q=2$) this term vanishes,
and the leading term is $\phi^4$, recovering the standard field
theory representation.  For $q\ne 2$ we can write
\begin{equation}
S=\int \d^d x \left[ \frac{1}{2} (\nabla \phi_i) (\nabla \phi_i) 
        +\frac{1}{2} m_0^2 \phi_i  \phi_i 
        +\frac{1}{3!} g_0 d_{ijk} \phi_i \phi_j \phi_k \right],
\label{ACTION3}
\end{equation}
where  the coefficients $d_{ijk}$ depend on the model
(Potts, percolation, Lee-Yang singularities, etc.), and $n \equiv q-1$ is the
number of components of the field $\phi_i$. Thus, the percolation is
described by the action (\ref{ACTION3}) in the limit of zero
components of the fields.

Using the standard tools it is possible to obtain an
$\epsilon$-expansion for this model (and in particular for the
percolation). The power counting tells us that the upper critical
dimension of the model is six and thereby the expansion parameter is
$\epsilon=6-d$. Results up to
three loops can be found in the literature \cite{ALCANTARA}.

For large dimensions ($d=5$, and, of course, $6$) there is a
good agreement between the results obtained from the
$\epsilon$-expansion (resumed using Pad\'e techniques), the values from
numerical simulations, and the results from high temperature expansions.

In lower dimensions, the results disagree for the anomalous dimension, $\eta$.
The $\epsilon$-expansion predicts a clear negative value, 
while in the two dimensional case  $\eta$ should be
non-negative because the
correlation function is decreasing with the distance.
In fact, in this case, it has been conjectured  \cite{NIENHUIS} that 
$\eta=5/24$.

In this paper we will show that the value of the four dimensional 
$\eta$ exponent turns
out to be large by a $30\%$, compared to the $\epsilon$-expansion.
Thereby it remains as an open problem to understand
why the convergence of the $\epsilon$-expansion for this model is so
poor even for small values of $\epsilon$~\cite{PAFU}. In order to calculate
critical exponents we extend some recently developed accurate finite-size
scaling techniques~\cite{OURFSS} to site percolation.
As a benchmark we report the two dimensional critical exponents (for which
there are almost exact analytical estimates).

A related model with the site percolation is the diluted Ising model
\cite{PARU}. It is defined as a standard Ising model where the spins
live only on filled (with probability $p$) sites.  The field
theoretical description of this model is a $\phi^4$-theory with a
random mass term. Using the replica trick it can be related with
an O($N$) symmetric $\phi^4$ theory with cubic anisotropy, in the limit
of zero field components (i.e. $N\rightarrow 0$) \cite{AMIT,PARISI}.

The limit of zero temperature (large $\beta$) of the diluted Ising
model is the site percolation while when $p\rightarrow 1$ it is the
pure Ising model. A precise determination of the critical exponents of
the $d=4$ percolation is also a very useful first step 
to understand the phase
diagram $(\beta,p)$ in the diluted Ising model. On the other hand, the site
percolation is useful as a benchmark to develop and test different
tools to apply to more complicated systems as the $d=4$ diluted
Ising model \cite{FUTURE}.

Finally, we remark that we are specially interested in these four
dimensional  models in
relation with the triviality issue (is there an interacting continuum
limit in four dimensions?).  In order to solve the triviality problem
is crucial to characterize all the possible fixed points in four
dimensions. The site percolation has the unusual feature of having the
critical dimension at $d=6$, thus, it does not present the usual
Mean Field exponents at $d=4$.

\section{\protect\label{S_NM}Numerical Methods}

We will work in a hypercubic lattice of linear size $L$ with periodic
boundary conditions.  The Monte Carlo (MC) procedure for
generating configurations in this model is straightforward:
we fill each lattice site with probability $p$. The next step is to
build the clusters, what is a deterministic procedure. To save
computer memory in the larger lattices, we use a self-recurrent
algorithm (in C language). In this way the total memory employed to
sketch the clusters is almost negligible (it grows nearly as the
lattice size squared).

Due to the absence of MC dynamics, the system is specially vulnerable
to eventual pathologies of the random number generator. We have
observed significant deviations in some quantities for a commonly used
shift register generator \cite{PARISI-RAPUANO}, specially in the
larger lattices. To avoid these effects, we have used as generator the
sum (modulus 1) of the output of the generator of ref.
\cite{PARISI-RAPUANO} and a congruential one, since it is known that
their respective drawbacks are very different.\footnote{
We have used \hbox{$X_{n+1}= 16807 X_n {\mod( 2^{31}-1)}$} for the
congruential random generator, whereas
the shift register formulas read: 
$X_n= X_{n-24}+X_{n-55}$; using as pseudorandom number 
$X_n$~{\sc xor}$~X_{n-61}$.}

To define the observables that we measure, it is useful to consider a
related model that is a diluted Ising model with nearest neighbors
infinite coupling, where the spins, $\sigma_i=\pm 1$, live only in
filled sites. It is easy to show that the magnetization of the latter
model,
\begin{equation}
{\cal M}=\frac{1}{V}\sum_i \sigma_i,
\end{equation}
$V$ being the volume,
coincides with the strength of the percolating cluster in the
thermodynamical limit and at $T=0$.

Knowing the size of the clusters, as their spins must take the same
sign, we can write
\begin{equation}
{\cal M}=\frac{1}{V}\sum_\mathrm{c} s_\mathrm{c}n_\mathrm{c},
\end{equation}
where $s_\mathrm{c}$ is the sign of the cluster c, $n_\mathrm{c}$
its size, and the sum runs over all clusters. As $s_\mathrm{c}$ are
statistically independent, 
we can construct an
improved estimator for even powers of  $\cal M$ (the only
non-vanishing in a finite lattice) averaging over all possible
values of $\{s_\mathrm{c}\}$, that henceforth we will denote as 
$\overline{(\cdot \cdot \cdot)}$. For the second power we have
\begin{equation}
\overline {{\cal M}^2}=\frac{1}{V^2}\sum_\mathrm{c}n_\mathrm{c}^2.
\end{equation}

We define the susceptibility as
\begin{equation}
\chi=V\left\langle \overline{{\cal M}^2} \right\rangle.
\end{equation}
To compute the Binder parameter $V_M$ 
we can construct an improved estimator for the fourth power of the
magnetization. Averaging over signs, we obtain after some algebra
\begin{equation}
\overline {{\cal M}^4}=
3\left(\overline {{\cal M}^2}\right)^2
-\frac{2}{V^4}\sum_\mathrm{c}n_\mathrm{c}^4,
\end{equation}
from which \footnote{For another application of the Binder cumulant
in percolation theory see \cite{HARRIS}.}
\begin{equation}
V_M=\frac{3}{2}-\frac{1}{2}\frac{\langle\overline {{\cal M}^4}\rangle}
           {\langle\overline {{\cal M}^2}\rangle^2}.
\end{equation}

For the finite-size scaling (FSS) method that we employ, it is very
useful an accurate measure of the correlation length. We have used
the second momentum definition~\cite{XIL} in the associated Ising model, that,
in a finite lattice, reads
\begin{equation}
\xi=\left(\frac{\chi/F-1}{4\sin^2(\pi/L)}\right)^{1/2},
\label{XI}
\end{equation}
where $F$ is defined in terms of the Fourier transform of the
magnetization
\begin{equation}
\widehat{\cal M}(\mathitbf{k})=\frac{1}{V}\sum_{\mathitbf{r}}\e^{\mathrm i
\mathitbf{k}\cdot\mathitbf{r}} \sigma_{\mathitbf r},
\end{equation}
as
\begin{equation}
F=\frac{1}{4}\left\langle |\widehat{\cal
M}(2\pi/L,0,0,0)|^2+\mathrm{permutations}\right\rangle.
\end{equation}

It is also possible to construct an improved estimator for $|\widehat
{\cal M}|^2$ as
\begin{equation}
\overline{|\widehat {\cal M}(\mathitbf{k})|^2}=\sum_\mathrm{c}|\widehat
n_\mathrm{c}(\mathitbf{k})|^2,\qquad \widehat n_\mathrm{c}(\mathitbf
k)\equiv\frac{1}{V}\sum_{\mathitbf r\in \mathrm c}
\e^{\mathrm i \mathitbf k\cdot\mathitbf r}.
\end{equation}

To measure the critical exponents we use a form of the FSS ansatz that
only involves measures on a finite lattice. For an operator $O$ that
diverges as $(p-p_\mathrm{c})^{-x_O}$, its mean value in a size $L$
lattice can be written, in the critical region, as
\begin{equation}
O(L,p)=L^{x_O/\nu}\left(F_O(\xi(L,p)/L)+O(L^{-\omega})\right), \label{FSS}
\end{equation}
where $F_O$ is a scaling function and $\omega$ is the
universal leading corrections-to-scaling exponent. 
From a Renormalization Group point of
view, $\omega$ corresponds to the leading irrelevant operator.

We can eliminate the unknown scaling
function using the values from two different lattice sizes measuring
at a $p$ value where the $\xi/L$ quotients match. Specifically, defining
\begin{equation}
Q_O=O(sL,p)/O(L,p),
\end{equation}
we can write
\begin{equation}
\left.Q_O\right|_{Q_\xi=s}=s^{x_O/\nu}+O(L^{-\omega}).\label{QUO}
\end{equation}
Other examples of application of this method can be found in
refs.~\cite{OURFSS}.

The form of the scaling corrections allows to parameterize the
finite-size effect on the determination of the critical exponents
as
\begin{equation}
\left(\frac{x_O}{\nu}\right)_\infty- \left(\frac{x_O}{\nu}\right)_{(L,sL)}
\propto L^{-\omega}\label{INFTYEXP}.
\end{equation}

To compute the $\omega$ exponent, we can use equation (\ref{FSS}) for
an operator with $x_O=0$ (as, for instance, $V_M$ or $\xi/L$)
obtaining for the shift of the crossing point of lattice sizes
$L$ and $sL$~\cite{BINDER}

\begin{equation}
\Delta p^{L,sL}\equiv \left[p_c(L,sL)-p_c(\infty)\right] \propto
\frac{1-s^{-\omega}}{s^{\frac{1}{\nu}}-1}L^{-\omega-\frac{1}{\nu}}.
\label{SHIFTBETA}
\end{equation}

To efficiently use  the {\sc FSS} formulas, it is necessary to use a
reweighting method to move in the critical region. For this
model there is not a Boltzmann weight, but the role of the energy
is carried out by the density of the configuration, and the
probability distribution is binomial.

The probability of finding a density $q$ when filling sites with a
probability $p$ is
\begin{equation}
\rho_p(q)=\frac{V!}{(qV)!((1-q)V)!}p^{qV}(1-p)^{(1-q)V}\label{BINOMIAL} .
\end{equation}
From a set of $N$ measures of an observable $O$ and the actual density
of the configuration $\{(O_i,q_i)\}$ we can compute the mean value
of the observable for a neighbor density $p'$ as
\begin{equation}
O(p')=\frac{1}{N}{\displaystyle\sum_i \frac{\rho_{p'}(q_i)}{\rho_p(q_i)}O_i}
     =\frac{1}{N}{\displaystyle\sum_i \left(\frac{p'}{p}\right)^{q_iV}
        \left(\frac{1-p'}{1-p}\right)^{(1-q_i)V}O_i}.\label{REWEIG} 
\end{equation}
Using equation (\ref{REWEIG}) $p$-derivatives of observables can also
be computed.

Obviously we cannot extrapolate much further than $\sqrt{p(1-p)/V}$,
which is the dispersion of the distribution~(\ref{BINOMIAL}).
Therefore the visible region decreases as
$L^{-d/2}$. Fortunately, it is enough for our purposes
since to use eq. (\ref{QUO}) we need to move in a
neighborhood of the critical point whose size decreases as
$L^{-\omega-1/\nu}(\approx L^{-2.5})$.

\section{\protect\label{S_NR}Numerical Results}

We have produced a million of independent samples for each 
$L^4$ lattices, with $L=8,12,16,24,32$ and 48. 

To measure the thermal critical exponent we have used as operators:
$\d\log\chi/\d p$ ($x_{\d\log\chi/\d p}=1$)
and $\d\xi/\d p$ ($x_{\d\xi/\d
p}=1+\nu$). For the magnetic exponents we have used the susceptibility
$\chi$ ($x_\chi=\gamma$).  We remark that, although $\chi$ is a fast varying
function of $p$ at the critical region (see refs.~\cite{OURFSS}), 
the use of eq. (\ref{QUO}) allows a very precise measure.
Moreover as what we directly measure is the quotient
$\gamma/\nu=2-\eta$, we can obtain a very accurate determination of the
anomalous dimension $\eta$.

We have checked the method in the $d=2$ case, where there is a very solid 
conjecture~\cite{NIENHUIS} for the values of the critical exponents,
which is confirmed by conformal group analysis.
We present the measured critical exponents for the two dimensional
site percolation in table~\ref{EXPO2D}, obtained from a million of
samples for each lattice size. The conjectured values by Nienhuis
\cite{NIENHUIS} are $\eta=5/24=0.20833\ldots$, $\nu=4/3$ and
$\omega=2$. The agreement is very good.

\begin{table}[t]
\begin{center}
\begin{tabular}{|r|l|l|l|}\hline
    & \multicolumn{2}{c|}{$\nu$} 
    & \multicolumn{1}{c|}{$\eta$} \\
\cline{2-3} \cline{3-4}
$L$ & \multicolumn{1}{|c|}{$\d\xi/\d p$} 
    & \multicolumn{1}{c|}{$\d \log (\chi)/\d p$} 
    & \multicolumn{1}{c|}{$\chi$} \\\hline\hline
24&1.324(9)&1.326(14)&0.2155(5)\\\hline
32&1.330(8)&1.30(2)  &0.2121(4)\\\hline 
48&1.344(10)&1.36(2)  &0.2085(4)\\\hline
64&1.330(9)&1.36(2)  &0.2082(4)\\\hline
\end{tabular}
\caption{Estimates for the critical exponents of two dimensional
site percolation obtained from the finite-size scaling analysis using
data from lattice sizes $L$ and $2L$.  In the second row we show
the operator used for each column.  Practically we can read from the
last row the conjectured values.}
\label{EXPO2D}
\end{center}
\end{table}

In the four dimensional case (see table \ref{EXPO4D}), we observe a
very stable value for the $\nu$ exponent when using the operator $\d
\xi/\d p$. However, the results for the exponents $\eta$ or $\nu$
computed from measures of other operators do need an infinite volume
extrapolation, what will be considered next.

To measure the critical density and the corrections-to-scaling
exponent $\omega$,
we have studied the crossing points of $V_M$ and $\xi/L$ for
different pairs of lattice sizes, fitting the displacements to the
functional form~(\ref{SHIFTBETA}). As the behavior of $V_M$ and
$\xi/L$ is very different regarding the corrections-to-scaling, we
obtain a great improvement performing a joint fit.

\begin{figure}[t]
\begin{center}
\leavevmode
\rotate[r]{\centering\epsfig{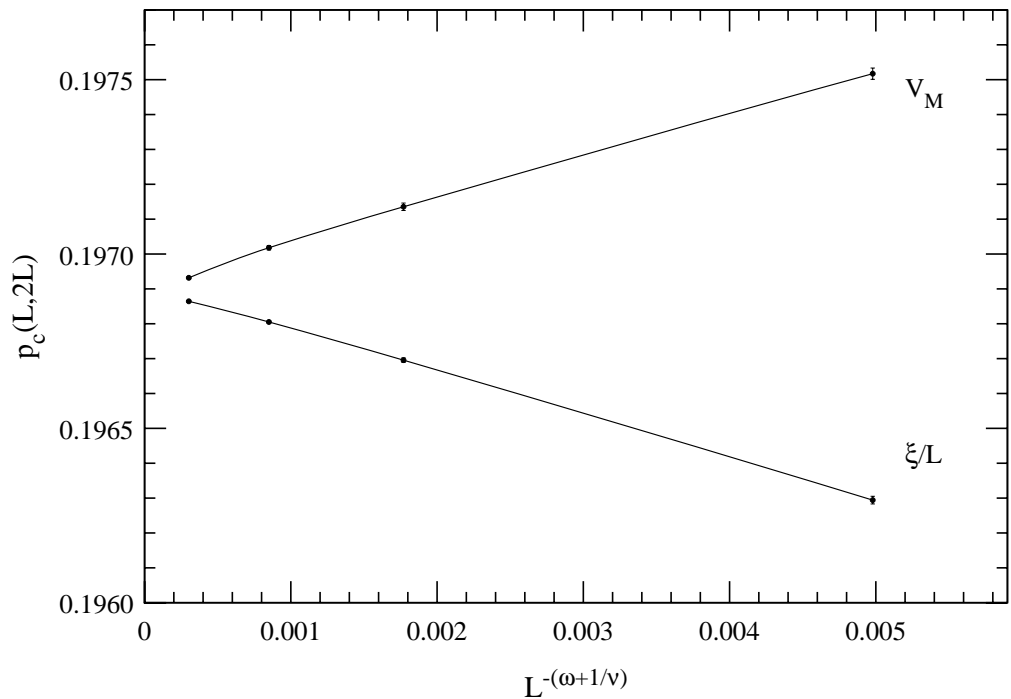}}
\end{center}
\protect\caption{$p_c(L,2L)$ as a function of
$L^{-(\omega+1/\nu)}$ for the observables $V_M$ and $\xi/L$.
\protect\label{F_V_M}}
\end{figure}

We show in figure \ref{F_V_M} the crossing points of $V_M$ and
$\xi/L$ as a function of $L^{-(\omega+1/\nu)}$, where we have used
$\nu=0.689$ and $\omega=1.13$.

We fix the lattices ratio to $s=2$ and perform the fit twice, for 
$L\geq 8$ and for $L\geq12$. In both cases we obtain compatible
values for the $\omega$ exponent and for the critical density. We get
acceptable fits, for example $\chi^2/\mathrm{d.o.f.}=4.7/4$ for the former.
We give the central values from the
former fit but with the error bars coming from the latter fit:
\begin{equation}
\omega=1.13(10),\quad p_c(\infty)=0.196901(5)\label{PCRITOMEGA} .
\end{equation}
The error bars have been slightly (20\%) increased to take into
account the error in the value of $\nu$.

\restylefloat{table}
\begin{table}[H]
\begin{center}
\begin{tabular}{|r|l|l|l|}\hline
    & \multicolumn{2}{c|}{$\nu$} 
    & \multicolumn{1}{c|}{$\eta$} \\
\cline{2-3} \cline{3-4}
$L$ & \multicolumn{1}{|c|}{$\d\xi/\d p$} 
    & \multicolumn{1}{c|}{$\d \log (\chi)/\d p$} 
    & \multicolumn{1}{c|}{$\chi$} \\\hline\hline
8 &      0.689(3)&        0.668(3)&      -0.0687(7)\\\hline
12&      0.687(3)&        0.666(4)&      -0.0775(7)\\\hline
16&      0.688(4)&        0.681(5)&      -0.0823(6)\\\hline
24&      0.691(5)&        0.683(6)&      -0.0868(8)\\\hline
\hline
$\infty$&0.689(10) &        0.683(12) &      -0.0944(17+11)\\\hline
\end{tabular}
\caption{Critical exponents obtained 
using data from lattice sizes $L$ and $2L$ for the
four dimensional site percolation. In the second
row we show the operator used for each column. The last
row corresponds to the infinite volume extrapolation using
(\ref{INFTYEXP}).}
\label{EXPO4D}
\end{center}
\end{table}

Using these values, we can obtain an infinite volume
extrapolation for the critical exponents by means of (\ref{INFTYEXP}).
To control that higher order scaling-corrections can be neglected,
we use an objective criterium.
We perform the fit considering data from lattices of sizes 
$L\geq L_{\mathrm{min}}$ and then repeat it discarding
the smallest lattices data. If both fits parameters (extrapolated value and
slope) are compatible, we keep the central values from the former fit and
error bars from the latter. We have found that $L_{\mathrm{min}}=8$ is enough
for our data.

\begin{figure}[t]
\begin{center}
\leavevmode
\rotate[r]{\centering\epsfig{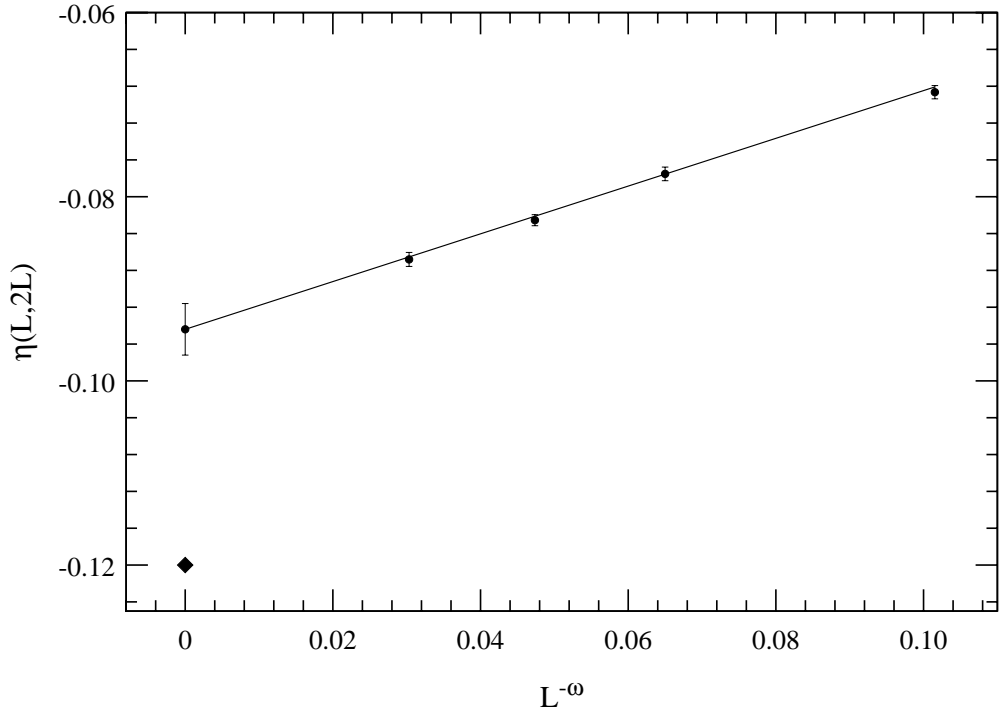}}
\end{center}
\protect\caption{$\eta(L,2L)$ as a function of $L^{-\omega}$, where
$\omega$ is the corrections-to-scaling exponent, that we have fixed in
this plot to $1.13$ (see text for more details). The numerical value
found in the literature, $\eta=-0.12$, is also displayed.\protect\label{F_eta}}
\end{figure}

The results are displayed
in the last row of table \ref{EXPO4D}. For $\eta$ the first term in the
error have been obtained considering $\omega$ fixed, and the second one
corresponds to the variation when $\omega$ moves within its error
bars. In figure \ref{F_eta} we show the behavior of $\eta(L,2L)$ as
a function of $L^{-\omega}$, with $\omega=1.13$, together with the 
extrapolated value.

\begin{table}[t]
\centering
\begin{tabular}{|c||c||c|c|c|} \hline
$d$ &  $\omega$  & $\nu$ & $\gamma$ &   $\eta$   \\ \hline \hline
6   &    0       & 0.5   &  1       &   0        \\ \hline
5   &    0.79    & 0.57  &  1.18    &   -0.07    \\ \hline
4   &    1.52    & 0.68  &  1.44    &   -0.12    \\ \hline    
3   &    2.23    & 0.83  &  1.81    &   -0.18    \\ \hline
2   &    2.95    & 1.07  &  2.41    &   -0.25    \\ \hline
\end{tabular}
\caption{ Our results using [2,1]-Pad\'e resummation for the $\omega$
exponent. We also report (columns three to five) the results for the
critical exponents ($\nu$, $\gamma$ and $\eta\equiv2-\gamma/\nu$)
obtained in ref.~\cite{ALCANTARA} using the [2,1]--Pad\'e-Borel
resummation.  }
\end{table}

At this point we can use the results of ref.~\cite{ALCANTARA}
obtained with the $\epsilon$-expansion.  We are specially interested
in the corrections-to-scaling exponent, that is, the derivate of
the $\beta$-function at the non trivial fixed point (i.e. $g^*
\neq 0$). Using the $\beta$-function and the non trivial fixed point
from ref.~\cite{ALCANTARA} we have obtained
\begin{equation}
\omega\equiv \left. \frac{\d \beta(g)}{\d g}\right|_{g^*}=
\epsilon-0.760767 \epsilon^2+2.00886\epsilon^3+ O(\epsilon^4).
\end{equation}
As $\epsilon(=6-d)$ is large, we have analyzed this series using the
Pad\'e technique.  Only the [2,1]--Pad\'e gives consistent results
(i.e. $\omega=O(\epsilon) >0$).  This agrees with the results of
ref.~\cite{ALCANTARA} where in the final Pad\'e analysis of their
series for the critical exponents only the [2,1]--Pad\'e is reported
(the results of the other Pad\'es turned out to be incompatible with
the numerical simulation results).  We show our results for $\omega$
in table 3. We also display in this table the results for the
exponents $\nu$, $\gamma$ and $\eta$ calculated in
ref.~\cite{ALCANTARA} using the [2,1]--Pad\'e-Borel resummation.

In other cases with $\epsilon=2$ 
a good agreement has been found  between resumed
series and numerical results. For the two dimensional Ising model the 
differences in $\eta$ and $\omega$  are
$1\%$ and $5-35\%$ respectively (taking into account the error bars
of the $\epsilon$-expansion estimate of $\omega$)\cite{ZINN}. However,
we have obtained a discrepancy of $30\%$ in
the anomalous dimension and of $50\%$ in the
$\omega$-exponent. Linking this discrepancy with the behavior of
$\eta$ with the dimension, reported in the introduction, we find
the $\epsilon$-expansion not trustworthy in this case.

\section{\protect\label{S_CON}Conclusions}

Using FSS techniques we have obtained accurate values for the critical
exponents of the four dimensional site percolation.  We have been able
to parameterize the leading corrections-to-scaling what allows to
largely reduce the systematic errors coming from finite-size effects.

We have obtained an anomalous dimension that is $30\%$ far away from 
previous numerical and analytical ($\epsilon$-expansion)
approaches.

We project to extend these methods to the case of the diluted Ising
model in four dimensions, in order to study the possible variation of
the critical exponents on the critical line \cite{FUTURE}.

\section*{Acknowledgments}

We thank to the CICyT (contracts AEN94-0218, AEN96-1634) for partial
financial support, specially for the financiation of dedicated Pentium
Pro machines where most of the simulations have been done.  JJRL is
supported by an EC HMC (ERBFMBICT950429) grant.

\hfill

\end{document}